\documentclass{iau}
\usepackage{graphicx,natbib}
 
\newcommand{\apj}{ApJ}           % Astrophysical Journal
\newcommand{\apjl}{ApJ}           % Astrophysical Journal
\newcommand{\mnras}{MNRAS}       % Monthly Notices of the RAS
\newcommand{\nat}{Nature}

\newcommand {\kms} {\ifmmode  \,\rm km\,s^{-1} \else $\,\rm km\,s^{-1}$ \fi }
\newcommand {\kpc} {\ifmmode  {\rm kpc}  \else ${\rm  kpc}$ \fi  }  
\newcommand {\Msun} {\ifmmode {M_{\odot}} \else ${M_{\odot}}$ \fi} 
\newcommand{\simf}{x} % IMF slope
\newcommand{\cg}{c_{-2}} 
\newcommand{\qh}{q_{\rm h}}
\newcommand{\vvir}{v_{\mathrm{vir}}} %% virial velocity
\newcommand{\Ms}{M_{\star}}
\newcommand{\slope}{\gamma}
\newcommand{\Reff}{R_{\mathrm{e}}}

\newcommand{\fDM}{f_{\mathrm{DM}}}

\title[3D structure of lens ETGs]{Dissecting the 3D structure of
  elliptical galaxies with gravitational lensing\\ and stellar
  kinematics}

\author[Barnab\`e, Spiniello \& Koopmans] {Matteo~Barnab\`e$^{1,2}$,
  Chiara~Spiniello$^3$ and L\'eon~V.~E.~Koopmans$^{4}$}

\affiliation{$^1$Dark Cosmology Centre, Niels Bohr Institute,
    University of Copenhagen,\\
    Juliane Maries Vej 30, 2100 Copenhagen {\O}, Denmark \\
    email: {\tt mbarnabe@dark-cosmology.dk} \\
  $^2$Niels Bohr International Academy, Niels Bohr Institute,
    University of Copenhagen, Blegdamsvej 17, 
    2100 Copenhagen {\O}, Denmark \\
    $^3$Max Planck Institute for Astrophysics,
    Karl-Schwarzschild-Strasse 1,\\ 85740 Garching, Germany \\
  $^4$Kapteyn Astronomical Institute, University of Groningen,
    P.~O.~Box 800,\\ 9700 AV Groningen, The Netherlands}

\pubyear{2014}
\volume{309}
\jname{Galaxies in 3D across the Universe}
\editors{B. L. Ziegler, F. Combes, H. Dannerbauer, M. Verdugo, eds.}

\begin{document}

\maketitle

\begin{abstract}
  The combination of strong gravitational lensing and stellar
  kinematics provides a powerful and robust method to investigate the
  mass and dynamical structure of early-type galaxies. We demonstrate
  this approach by analysing two massive ellipticals from the XLENS
  Survey for which both high-resolution \emph{HST} imaging and
  X-Shooter spectroscopic observations are available. We adopt a
  flexible axisymmetric two-component mass model for the lens
  galaxies, consisting of a generalised NFW dark halo and a realistic
  self-gravitating stellar mass distribution. For both systems, we put
  constraints on the dark halo inner structure and flattening, and we
  find that they are dominated by the luminous component within one
  effective radius. By comparing the tight inferences on the stellar
  mass from the combined lensing and dynamics analysis with the values
  obtained from stellar population studies, we conclude that both
  galaxies are characterised by a Salpeter-like stellar initial mass
  function.  
\end{abstract}

\firstsection

%%%%%%%%%%%%%%%%%%%%%%%%%%% INTRODUCTION %%%%%%%%%%%%%%%%%%%%%%%%%%%% 

\section{Introduction}
\label{sec:introduction}

Understanding the formation and evolution mechanisms of early-type
galaxies (ETGs) remains a crucial challenge in present-day
astrophysics. In recent years, high-resolution simulations of galaxy
evolution, including both baryons and dark matter
\citep[e.g.][]{Genel2014}, have progressed enough that it is finally
becoming possible to investigate in detail the physical properties of
simulated systems and compare them to the corresponding properties of
observed ellipticals in the local Universe. Whereas nearby ETGs have
been the object of intense scrutiny, with stellar dynamics being the
most often employed diagnostic tool \citep[see e.g.][]{Thomas2011,
  Cappellari2013-atlasXX}, little is known about the detailed
structure of ellipticals beyond redshift $z \approx 0.1$, despite the
obvious importance of providing observational constraints to
simulations throughout cosmic history.

At higher redshift, however, massive galaxies occasionally act as
strong gravitational lenses: for such systems it becomes possible to
supplement stellar kinematics with the lensing constraints to
investigate the mass structure of these distant galaxies to a level
that is comparable with what can be obtained for nearby objects
\citep[see e.g.][]{Czoske2008, Koopmans2009, vandeVen2010,
  Barnabe2010, Barnabe2011}. High quality combined data-sets,
including extended kinematics, make it possible to disentangle the
contributions of dark and luminous matter to the total mass budget,
which in turns allows one to draw inferences also on the stellar
initial mass function (IMF) of the system, by comparing the results
with the stellar masses obtained from independent stellar population
synthesis (SPS) analysis assuming a variety of IMF profiles. We
demonstrate the power of this approach by applying a self-consistent
combined lensing and dynamics analysis to two massive ETGs from the
XLENS Survey \citep{Spiniello2011} for which high signal-to-noise
X-Shooter spectroscopic observations are at hand.

%%%%%%%%%%%%%%%%%%%%%%%%%%% OBSERVATIONS %%%%%%%%%%%%%%%%%%%%%%%%%%%% 

\section{Observations}
\label{sec:observations}

Both massive ellipticals SDSS\,J0936$+$0913 (velocity dispersion
$\sigma \simeq 245$ $\kms$; redshift $z = 0.164$) and
SDSS\,J0912$+$0029 ($\sigma \simeq 325$ $\kms$; $z = 0.190$) were
first observed as part of the SLACS Survey \citep{Bolton2008a}. The
high-resolution images needed for the lensing modelling were obtained
with \emph{HST} ACS through the \emph{F814W} filter. Elliptical
B-spline models of the lens galaxies were subtracted off the images to
isolate the structure of the gravitationally lensed background
sources.

High signal-to-noise ($\mathrm{S/N} > 50$) \emph{UVB}--\emph{VIS}
X-Shooter spectra are available for both systems, and were used both
to carry out SPS analysis and derive long-slit spatially resolved
stellar kinematics up to about one effective radius, $\Reff$.

%%%%%%%%%%%%%%%%%%%%%%%% MASS MODEL %%%%%%%%%%%%%%%%%%%%%%%%%%%%%%%%%

\section{Modelling}
\label{sec:mass-model}

A joint self-consistent modelling of all the available lensing and
kinematic constraints to derive inferences on the mass and dynamical
structure of the two systems is conducted with the fully Bayesian
\textsc{cauldron} code \citep{Barnabe-Koopmans2007, Barnabe2012}.  The
lens galaxy is described by a flexible and realistic two-component
axisymmetric mass model consisting of (i) a generalised
Navarro-Frenk-White (gNFW) halo characterised by four free parameters
(the inner density slope~$\slope$, the axial ratio~$\qh$, the halo
concentration parameter~$\cg$ and the virial velocity~$\vvir$) and
(ii) a mass profile for the baryonic component obtained by
de-projecting the multi-Gaussian expansion fit (MGE) to the observed
surface brightness distribution of the galaxy, characterised by one
free parameter (the total stellar mass~$\Ms$ that sets the
normalisation of the luminous distribution). The stellar kinematics is
modelled with the Jeans anisotropic method (JAM) developed by
\citet{Cappellari2008}, including the meridional plane orbital
anisotropy~$b = \sigma^{2}_{R}/\sigma^{2}_{z}$ as a free parameter.

%%%%%%%%%%%%%%%%%%%%%%%% RESULTS %%%%%%%%%%%%%%%%%%%%%

\section{Results and discussion}
\label{sec:results}

\begin{figure}

  \centering
  \resizebox{0.90\hsize}{!}{\includegraphics[angle=0]
    {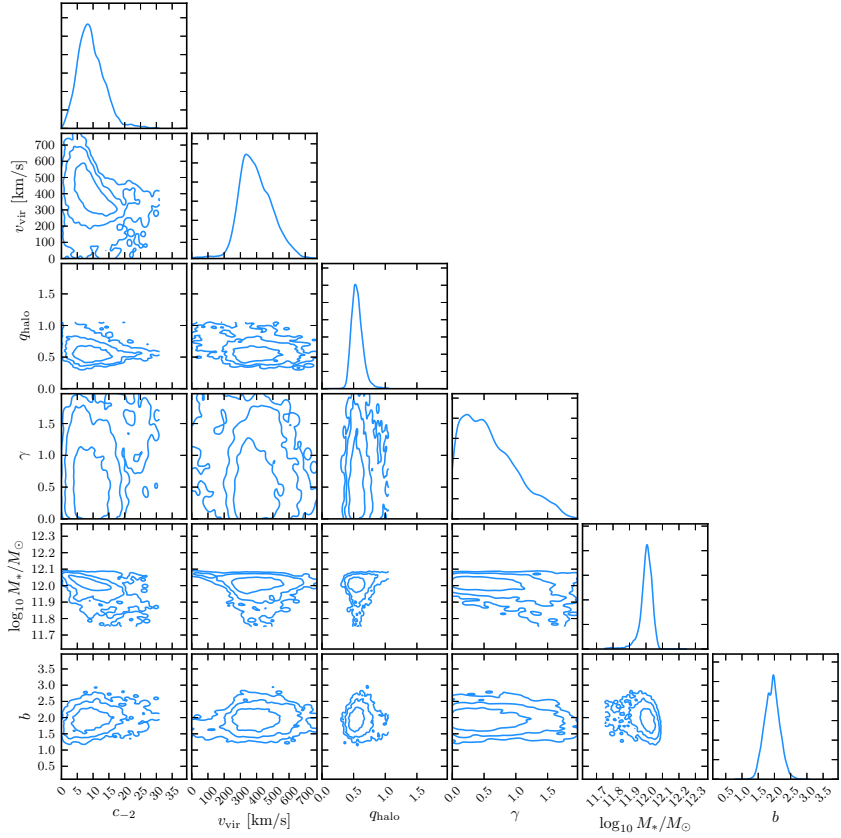}}
%  \put(-190,310){\Large lensing + kinematics}
%  \put(-190,290){\Large inferences for J0912}
  \caption{Marginalised 1D and 2D posterior PDFs for the six model
    parameters for galaxy J0912, using the combined constraints from
    lensing and kinematics. The three contours indicate the regions
    containing, respectively, 68\%, 95\% and 99.7\% of the
    probability.}
  \label{fig:cornerplot}

\end{figure}

The full inferences on the model parameters obtained from the combined
\textsc{cauldron} analysis are expressed as multivariate posterior
probability distribution functions (PDFs), encapsulating all the
information on the uncertainties in a statistically rigorous way.  The
1D and 2D marginalised posterior PDFs for the individual parameters
can be visualised as a familiar corner plot, as shown in
Fig.~\ref{fig:cornerplot} for galaxy J0912. 

% --------------------- CORNERPLOT: TOT -----------------------------
% -------------------------------------------------------------------

Both systems are found to be markedly dominated by the baryonic
component within the inner~$\Reff$, with a dark matter fraction $\fDM
= 0.06^{+0.10}_{-0.05}$ for J0936 and $\fDM = 0.18^{+0.08}_{-0.08}$
for the more massive system J0912. The dark halo structure of both
galaxies ($\cg = 6.3^{+18.4}_{-4.1}$ and $\vvir = 124^{+160}_{-88}$
$\kms$ for J0936; $\cg = 7.2^{+3.8}_{-2.7}$ and $\vvir =
470^{+160}_{-130}$ $\kms$ for J0912) is consistent with the
concentration--virial velocity relation predicted from pure $N$-body
simulations \citep{Maccio2008}: thus, within the fairly broad
uncertainties, there is no obvious evidence for contraction or
expansion of the dark halo in response to galaxy formation. The inner
slope of the dark halo is basically unconstrained for J0936 ($\slope =
0.92^{+0.72}_{-0.64}$), while for J0912 we can conclude that steep
slopes are disfavoured ($\slope = 0.46^{+0.41}_{-0.30}$). The two
haloes differ in terms of flattening, with J0936 having a more
spherical halo ($\qh = 0.93^{+0.25}_{-0.18}$), while the halo of J0912
is clearly oblate ($\qh = 0.54^{+0.10}_{-0.08}$).

The two lens galaxies have dissimilar dynamical structures within the
probed region: J0936 ($b = 0.89^{+0.37}_{-0.31}$) is consistent with
being a semi-isotropic system, while the velocity dispersion ellipsoid
of J0912 is distinctly flattened ($b = 1.94^{+0.24}_{-0.21}$).

% ----------------------- COMBINED INFERENCES -----------------------
\begin{figure}

  \centering
  \resizebox{0.83\hsize}{!}{\includegraphics[angle=-90]
    {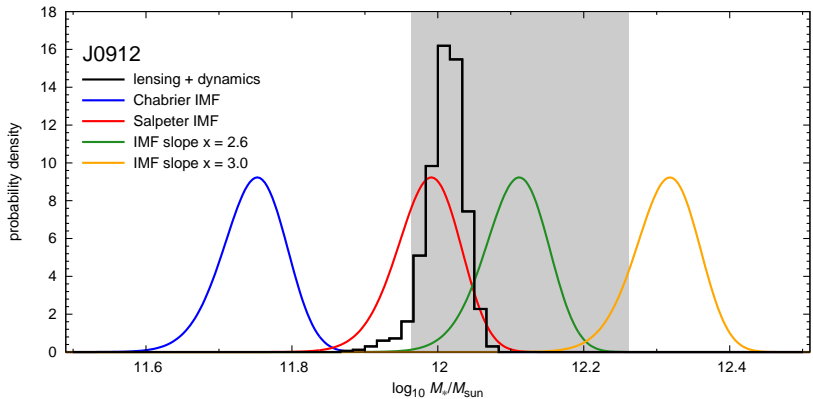}}
  \caption{Comparison of the total stellar mass inferred for J0912
    from the combined lensing and dynamics study (histogram)
    with~$M_{\star}$ obtained from SPS analysis, for different values
    of the IMF slope (coloured curves). The green curve indicates
    $M_{\star}$ for the best-fit IMF slope determined from an
    independent method based on spectroscopic simple stellar
    population modelling of line-strength indices, with the grey band
    showing the 1-sigma uncertainties around that value.}
 \label{fig:IMF}

\end{figure}
% -------------------------------------------------------------------

By comparing the stellar mass inferred from the combined lensing and
dynamics modelling with the one derived from SPS analysis of X-Shooter
spectra we can also put robust constraints on the normalisation of the
stellar IMF, as illustrated in Fig.~\ref{fig:IMF} for J0912. We
determine that both galaxies are consistent with having a Salpeter IMF
(i.e., slope $\simf = 2.35$ for a power-law IMF profile $dN/dm \propto
m^{-\simf}$), in agreement with the findings for local massive ETGs
from both spectroscopic and dynamical studies
\citep[e.g.][]{vanDokkum-Conroy2010, Cappellari2012}. Moreover, we can
rule out both a Chabrier and a ``super-Salpeter'' ($\simf \ge 3$) IMF
with a high degree of confidence. The XLENS data-set makes it possible
also to determine the IMF in a completely independent way from a
spectroscopic simple stellar population analysis of the optical
line-strength indices, using the approach outlined by
\citealt{Spiniello2014}: the IMF slopes obtained with this method
($\simf = 2.10 \pm 0.15$ for J0936 and $\simf = 2.60 \pm 0.30$ for
J0912) are fully consistent with the much tighter lensing and dynamics
results. Finally, the constraints from these two complementary
approaches can be combined to investigate the elusive low-mass cut-off
of the IMF, as demonstrated by \citet{Barnabe2013}.  Work is ongoing
to extend this analysis to the entire XLENS sample of~10 massive lens
ETGs.

%%%%%%%%%%%%%%%%%%%%%%%%%%% ACKNOWLEDGMENTS %%%%%%%%%%%%%%%%%%%%%%%%%

\section*{Acknowledgments}

The Dark Cosmology Centre is funded by the Danish National Research
Foundation.

%%%%%%%%%%%%%%%%%%%%%%%%%%%%% BIBLIOGRAPHY %%%%%%%%%%%%%%%%%%%%%%%%%%

\label{lastpage}

\end{document}